\input{aipcheck}

\documentclass[
    ,final            
  ]
  {aipproc}

\layoutstyle{8x11single}
\usepackage{amsmath}
\usepackage[abs]{overpic}
\usepackage{graphicx}
\usepackage{color}

\begin{document}

\title{Charge exchange reaction by Reggeon exchange and W$^{+}$W$^{-}$-fusion}


\author{R. Schicker}{
  address={Phys. Inst., University Heidelberg}
}

\begin{abstract} Charge exchange reactions at high energies are examined. 
The existing cross section data on the Reggeon induced reaction 
pp $\rightarrow$ n + $\Delta^{++}$ taken at the ZGS and ISR accelerators are 
extrapolated to the energies of the RHIC and LHC colliders. The interest in 
the charge exchange reaction induced by $W^{\pm}$-fusion is presented, and 
the corresponding QCD-background is examined.    
\end{abstract}

\maketitle


\section{Introduction}

The hadronic cross section is a fundamental quantity, its energy dependence can,
however, so far not be derived by first principles from the QCD-Lagrangian.
Within Regge phenomenology, the strong interaction is due to the exchange 
of Reggeons which are characterised by their trajectory. At high energies,
hadronic interactions are dominated by the exchange of an additional trajectory,
the Pomeron. The energy dependence of the total hadronic cross section reflects 
therefore the interplay between Reggeon and Pomeron contributions. 
A three component fit yields good agreement with the existing data 
both for proton-proton and proton-antiproton cross sections in the 
range from $\sqrt{s}$ = 23 GeV to  $\sqrt{s}$ = 8 TeV. 
Here, the three components are defined by the 
$f_{2},a_{2}$, the $\rho,\omega$ and the Pomeron-trajectory \cite{DL}. 
  
Hadronic charge exchange reactions can only be due to the exchange of 
charged Reggeons, and are therefore of interest for testing the Regge 
prediction of their energy behaviour. The transfer of electric charge
from one beam particle to a particle of the opposite beam is associated 
with bremsstrahlung radiation \cite{Jackson}. The theorem of Low relates 
the radiative leading and next-to-leading order in photon energy of the 
bremsstrahlung amplitude to the corresponding non-radiative 
amplitude \cite{Low}. The theorem of Low refers to soft photons,
i.e. to photon energies which are smaller than any other momentum scale 
in the process. A study of photon emission in the high energy limit finds 
corrections in the next-to-leading radiative amplitude due to the internal 
structure of the external states \cite{DelDuca}. The production of 
non-abelian  gauge bosons and gravitons is studied as a generalization 
of the theorem of Low in Ref. \cite{Lipatov}.

\section{Reggeon charge exchange reaction}

In hadronic charge exchange reactions in proton-proton collisions, one 
or both of the beam particles can be excited to an N$^{*}$ or a 
$\Delta$-resonance. Here, the N$^{*}$ and $\Delta$-resonance represent 
the family of N$^{*}$ and $\Delta$-resonances as listed by the 
Particle Data Group \cite{DPG}.

\begin{eqnarray}  
pp \hspace{0.4cm}\rightarrow & n + \Delta^{++} \hspace{0.4cm}\rightarrow & n + p\;\pi^{+} \\
pp \hspace{0.4cm}\rightarrow & \Delta^{0} + \Delta^{++} \hspace{0.4cm}\rightarrow & p\;\pi^{-} + p\;\pi^{+} \\
pp \hspace{0.4cm}\rightarrow & \Delta^{0} + \Delta^{++} \hspace{0.4cm}\rightarrow & n\;\pi^{0} + p\;\pi^{+} 
\end{eqnarray}  

The charge exchange reactions 1)-3) listed above can be measured by identifying
the forward scattered remnants of the resonance decay. In the following, only 
reaction 1) in the list above is discussed to illustrate the study of such 
reactions. The reactions 2) and 3) are the same, but are listed here separately
due to the different final state produced by the decay of the 
$\Delta^{0}$-resonance into charged and neutral mode, respectively. The 
identification of the different decay channels necessitates not only the 
measurements of forward charged tracks, but also of very forward focussed 
neutral particles like neutrons and $\pi^{0}$. 

\subsection{Cross section of Reggeon charge exchange reaction}

The reaction 1) listed above was measured in a spark chamber experiment at 
the Argonne National Zero Gradient Synchrotron (ZGS) at a beam momentum 
of $P_{lab}$ = 6 GeV/c. The analysis of the data found good agreement with 
the basic Chew-Low one-pion-exchange mechanism yielding an energy dependence 
$\propto P_{lab}^{-2}$ \cite{ZGS}. Later experiments with the Split Field 
Magnet (SFM) at the Intersecting Storage Ring (ISR) at CERN extended the 
cross section measurement of this reaction to the energy range of $\sqrt{s}$ 
= 23 to 53 GeV. The analysis of these data resulted in the conclusion that
$\rho, a_{2}$-exchange is the dominating mechanism at 
the ISR-energies \cite{ISR}. 

\begin{figure}[h]
\begin{minipage}[h]{.96\textwidth}
\vspace{.2cm} 
\begin{overpic}[width=.44\textwidth]{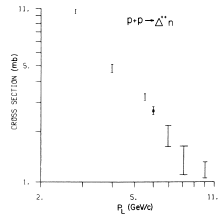}
\put(140.,160.){\it\footnotesize J.D. Mountz et al.}
\put(140.,150.){\it\footnotesize Phys.Rev.D}
\put(140.,140.){\it\footnotesize Vol.12,5(1975)1211}
\thicklines
\end{overpic}
\hspace{1.6cm}
\begin{overpic}[width=.44\textwidth]{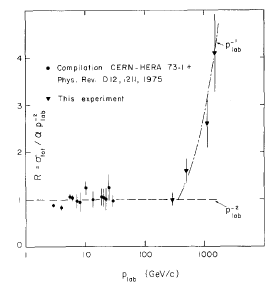}
\put(144.,90.){\it\footnotesize H. De Kerret et al.}
\put(144.,80.){\it\footnotesize Phys.Lett.B}
\put(144.,70.){\it\footnotesize Vol.69,3(1977)372}
\linethickness{.4mm}
\put(30.,48.){\vector(1.,0.) {84.}}
\put(50.,38.){\footnotesize $\pi$-exch.}
\put(120.,48.){\vector(1.,0.) {30.}}
\put(122.,38.){\footnotesize $\rho,a_2$-exch.}
\end{overpic}
\end{minipage}
\caption{Cross section data for pp $\rightarrow$ n $\Delta^{++}$ (Figure taken 
from Ref. \cite{ZGS}(left),  \cite{ISR}(right)).}
\label{fig1}
\end{figure}
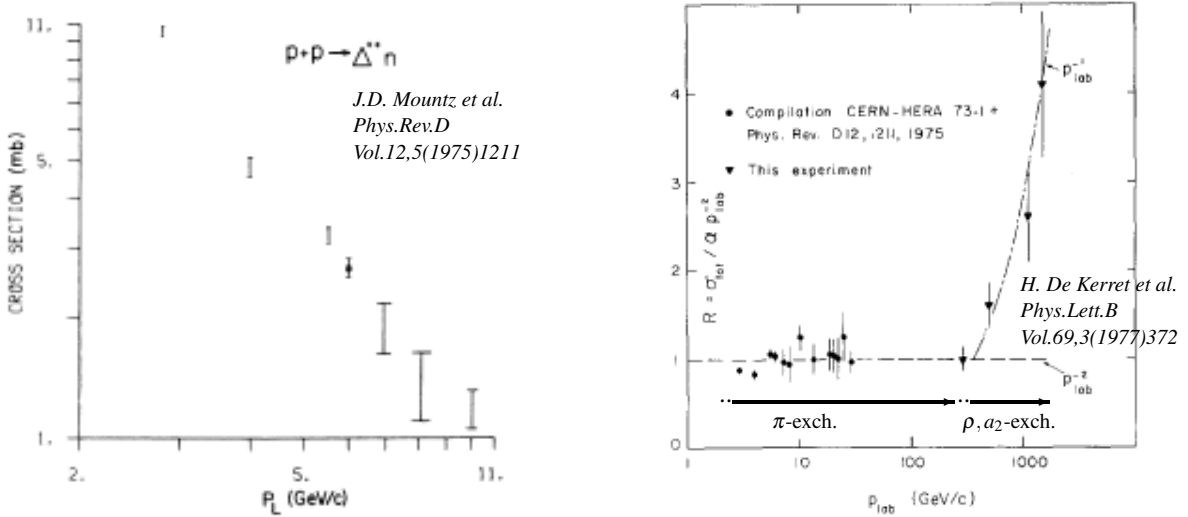
 
In Figure \ref{fig1}, the cross section for the charge exchange reaction
pp $\rightarrow$ n $\Delta^{++}$ is shown on the left for beam momenta $P_{lab} <$ 10 GeV/c. 
On the right hand side in Figure \ref{fig1}, the vertical axis is scaled with $P_{lab}^{-2}$. In this 
representation, the data points of pion-exchange dominated cross sections fall on a constant line.
Clearly visible is the transition from pion to $\rho, a_{2}$-dominated exchange at a beam momentum 
which corresponds approximately  to a centre-of-mass energy \mbox{$\sqrt{s} \sim$ 23 GeV,} i.e. 
the lowest energy at which data were taken at the ISR.

The cross section presented in Figure \ref{fig1} above can be extrapolated to energies beyond 
the ISR measurements with the assumption of momentum scaling $\propto P_{lab}^{-1}$ corresponding 
to $\rho, a_{2}$-exchange. 

\begin{table}[h]
\begin{tabular}{lccc}
\hline
  & \tablehead{1}{c}{b}{$\sqrt{\text{s}} \text{(GeV)}$}
  & \tablehead{1}{c}{b}{$\sigma$(nb)} \\
\hline
ISR & 31 & 580$\pm$90 & \\
& 45 & 210 $\pm$40 & \\
& 53 & 170 $\pm$40 & \\
\hline
RHIC & 100 & 48.5$\pm$5.5 \\
& 200 & 12.2$\pm$1.3 \\
\hline
LHC & 7x$10^3$ & (10.0$\pm$1.1)$\times 10^{-3}$ \\
& 14$\times 10^3$ & (2.4$\pm$0.3)$\times 10^{-3}$ \\
\hline
\end{tabular}
\caption{Cross section pp $\rightarrow$ n $\Delta^{++}$.}
\label{table1}
\end{table}
 
In Table \ref{table1}, the cross section values extrapolated to the RHIC and LHC energies are shown.
For this extrapolation, the cross section values at the ISR energies of $\sqrt{s}$ = 31, 45 and 53 GeV
and the quoted uncertainties were taken from \mbox{Ref. \cite{ISR},} and fitted with a 
$\propto P_{lab}^{-1}$\,-\,dependence.
The extrapolated cross section values are in the range of tens of nanobarns for the RHIC energies,
whereas the values at the LHC energies are on the order of a few picobarns. 

The characteristics of these events are very forward scattered beam remnants and a rapidity
gap in between, i.e. a region in rapidity where no particles are produced.
These events carry therefore the single gap topology of single and double-diffractive 
dissociation reactions. A double gap topology is also possible in hadronic charge
exchange reactions. Here, two charged Reggeons fuse and produce a hadronic
state at or close to mid-rapidity.  

\section{Charge exchange reaction by W$^{+}$,W$^{-}$ -\:exchange}

The double gap topology discussed above of forward beam remnants with activity 
in between can also be due to Weak-boson fusion (WBF) reactions. The study of such reactions is
of interest for gaining insight into the electroweak gauge boson sector of the Standard Model. 
Such WBF-reactions are sensitive to the electroweak symmetry breaking mechanism, and hence
are of high interest to improve the current understanding of this symmetry breaking. 

\begin{figure}[h]
\begin{minipage}[h]{.92\textwidth}
\vspace{.0cm} 
\begin{overpic}[width=.26\textwidth]{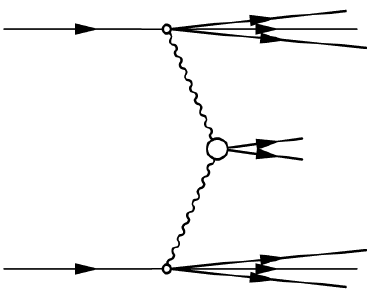}
\put(45.,25.){\footnotesize W$^{+}$}
\put(45.,56.){\footnotesize W$^{-}$}
\put(60.,-3.){\footnotesize p$^{*,0}$}
\put(60.,86.){\footnotesize p$^{*,2+}$}
\put(95.,42.){\footnotesize X}
\put(120.,42.){=}
\thicklines
\end{overpic}
\hspace{1.cm}
\begin{overpic}[width=.26\textwidth]{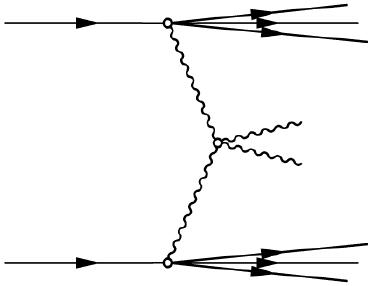}
\put(45.,25.){\footnotesize W$^{+}$}
\put(45.,56.){\footnotesize W$^{-}$}
\put(60.,-3.){\footnotesize p$^{*,0}$}
\put(60.,86.){\footnotesize p$^{*,2+}$}
\put(95.,34.){\footnotesize W$^{+}$\!,\,Z\,,$\gamma$}
\put(95.,50.){\footnotesize W$^{-}$\!,\,Z\,,$\gamma$}
\put(125.,6.){,}
\end{overpic}
\hspace{1.cm}
\begin{overpic}[width=.26\textwidth]{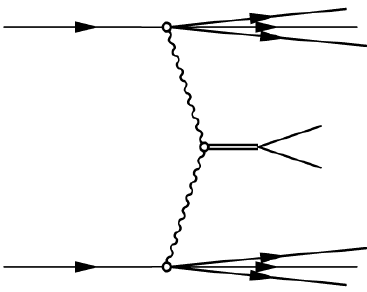}
\put(42.,25.){\footnotesize W$^{+}$}
\put(42.,56.){\footnotesize W$^{-}$}
\put(60.,-3.){\footnotesize p$^{*,0}$}
\put(60.,86.){\footnotesize p$^{*,2+}$}
\put(100.,42.){\footnotesize H}
\end{overpic}
\end{minipage}
\caption{W$^{+}$W$^{-}$\,-\,vector boson fusion.}
\label{fig2}
\end{figure}
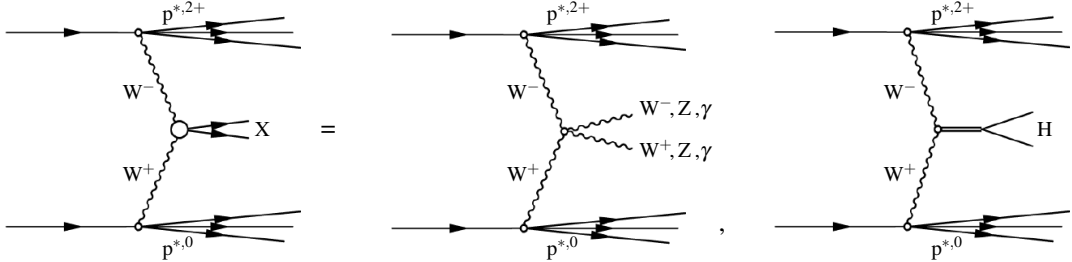 

In Figure \ref{fig2}, the vector boson fusion diagram W$^{+}$W$^{-} 
\rightarrow $X is shown on the left. In the middle, the diagram of interest 
for studies of quartic-couplings is displayed, whereas on the right, 
resonant WBF-production of Higgs is shown. A variety of different diagrams 
exist which can lead to the same final state. A careful study of these 
background channels is therefore mandatory to evaluate the signal to 
background ratio of WBF-measurements. 

\subsection{Background single quark exchange}

The background processes leading to the same final state as shown in Figure \ref{fig2}
consist of single and double quark exchange diagrams. The single quark exchange
can be written as $q\overline{q} \rightarrow$ W$^{+}$W$^{-}$\!, ZZ, $\gamma\gamma$ plus
additional parton exchange.

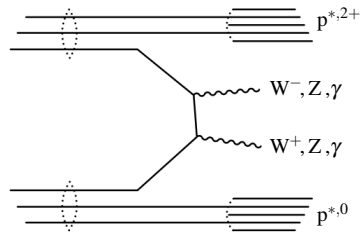
\begin{figure}[h]
\begin{minipage}[t]{.6\textwidth}
\begin{center}
\vspace{.0cm} 
\begin{overpic}[width=.42\textwidth]{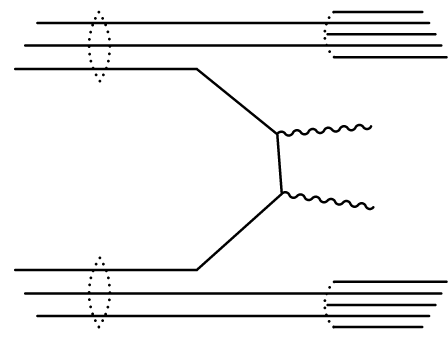}
\put(102.,32.){\footnotesize W$^{+}$\!,\,Z\,,$\gamma$}
\put(102,54.){\footnotesize W$^{-}$\!,\,Z\,,$\gamma$}
\put(120.,6.){\footnotesize p$^{*,0}$}
\put(120.,80.){\footnotesize p$^{*,2+}$}
\thicklines
\end{overpic}
\end{center}
\end{minipage}
\caption{Single quark exchange diagram.}
\label{fig3}
\end{figure} 

In Figure \ref{fig3}, the single quark exchange diagram is shown. A quark of one hadron
annihilates with an anti-quark from the sea-distribution of the other hadron and emits
a pair of W$^{+}$W$^{-}$, ZZ or photons. 

\subsection{Background double quark exchange}

The double quark exchange is expressed as 
$q\overline{q}q\overline{q} \rightarrow$ W$^{+}$W$^{-}$\!, ZZ, $\gamma\gamma$.

\begin{figure}[h]
\begin{minipage}[t]{.6\textwidth}
\begin{center}
\vspace{.0cm} 
\begin{overpic}[width=.42\textwidth]{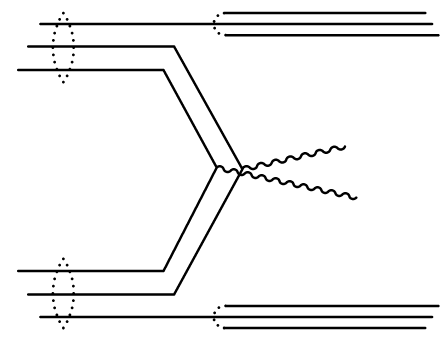}
\put(102.,32.){\footnotesize W$^{+}$\!,\,Z\,,$\gamma$}
\put(102.,54.){\footnotesize W$^{-}$\!,\,Z\,,$\gamma$}
\put(122.,6.){\footnotesize p$^{*,0}$}
\put(122.,85.){\footnotesize p$^{*,2+}$}
\thicklines
\end{overpic}
\end{center}
\end{minipage}
\caption{Double quark exchange diagram.}
\label{fig4}
\end{figure}
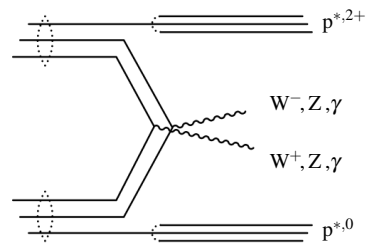

In the double quark exchange shown in Figure \ref{fig4}, two annihilation processes
of a quark of one hadron with an anti-quark of the other hadron generate the two
gauge bosons of the final state. In order to realistically evaluate the single and
double-quark exchange background, the cross section needs to be known differentially
as function of p$^{*,0}$ and p$^{*,2+}$, the invariant masses of the proton fragments.
In addition, the phase space distribution of these fragments needs to be known accurately 
in order to evaluate the experimental acceptance of the forward scattered proton remnants.

\section{Experimental considerations}

The experimental measurement  of the charge exchange reactions discussed above is in principle 
possible by identifying charge asymmetric forward systems with Z$_{tot}$ = 0  and Z$_{tot}$ = 2 at
the two opposite sides, respectively. The difference of these forward charges, $\Delta Z_{tot}$ = 2, 
constitutes a parameter for selecting the charge exchange events. Such measurements require, 
however, forward detector systems with excellent acceptance. The missing of one charged 
track would change $\Delta Z$ by one unit, but would still suffice to recognize a forward
charge asymmetry. The missing of two forward charged tracks would, however, introduce 
considerable uncertainty. If the two tracks are of like-sign charge and missed on opposite 
sides, or of unlike-sign charge and missed on the same side, then the charge asymmetry would 
not change. If the two tracks are of like-sign charge and missed on the same  side, or of 
unlike-sign charge and missed on opposites  sides, then the charge asymmetry would change by 
two units with resulting  values of Z$_{tot}$ = 0 or 4, respectively. 
The building and installation of forward detector systems with close to perfect acceptance
for charged tracks is, however, a considerable challenge from the experimental viewpoint.
Such systems are presently not in operation or under discussion neither at RHIC nor at the LHC.    

\section{Summary}

The cross section of the charge exchange reaction pp 
$\rightarrow$ n $\Delta^{++}$ is expected to follow an approximate  
$\propto p_{lab}^{-1}$-dependence at high energies as seen in the data taken at 
the ISR-energies. The precise measurement of this cross section at energies 
beyond the ISR energies would  allow to test the individual contributions 
from $\rho$ and $a_2$-exchanges due to their different energy dependence. 
The study of charge exchange by WBF-reactions necessitates a careful 
evaluation of the signal and of the background resulting from single and 
double-quark exchange diagrams. In order to optimize the signal to background 
ratio, the phase space distribution of the forward proton fragments needs 
to be known both for signal and background. The forward charge asymmetry of 
charge exchange reactions defines a parameter which can be used to select 
such events. The accurate  measurement of this charge asymmetry  poses, 
however, a considerable challenge to the experimenters, and has so far 
not been considered. 

\begin{theacknowledgments}
The author gratefully acknowledges fruitful discussions with J. Bartels,
L. Jenkovszky, L. Lipatov and R. Pasechnik on the issues presented here.
This work is supported by the German Federal Ministry of Education and 
Research under promotional reference 05P12VHCA1 and by WP8 of the hadron 
physics programme of the 7th EU programme period.
\end{theacknowledgments}



\bibliographystyle{aipproc}   




\end{document}